\documentclass[twocolumn,showpacs,preprintnumbers,amsmath,amssymb]{revtex4-1}

\usepackage{graphicx}
\usepackage{bm}
\usepackage{amsmath}

\begin{document}
 
\title{Viscoelasticity of two-layer-vesicles in solution}
 
\author{C.-Y. David Lu}
\email{cydlu@ntu.edu.tw}
\affiliation{Department of Chemistry and Department of Physics, Center of
Theoretical Physics, National Taiwan University, Taipei 106,
Taiwan}

\author{Shigeyuki Komura}
\affiliation{Department of Chemistry, Graduate School of Science and Engineering, 
Tokyo Metropolitan University, Tokyo 192-0397, Japan}

\author{Kazuhiko Seki}
\affiliation{National Institute of Advanced Industrial Science and Technology (AIST), 
Tsukuba 305-8565, Japan}

\date{\today}

\begin{abstract}
The dynamic shape relaxation of the two-layer-vesicle is calculated. 
In additional to the undulation relaxation where the two bilayers move in 
the same direction, the squeezing mode appears when the gap between the 
two bilayers is small. 
At large gap, the inner vesicle relaxes much faster, whereas the slow mode 
is mainly due to the outer layer relaxation.  
We have calculated the viscoelasticity of the dilute two-layer-vesicle suspension. 
It is found that for small gap, the applied shear drives the undulation mode 
strongly while the slow squeezing mode is not much excited. 
In this limit the complex viscosity is dominated by the fast mode contribution.
On the other hand, the slow mode is strongly driven by shear for larger gap.
We have determined the crossover gap which depends on the interaction between 
the two bilayers.  
For a series of samples where the gap is changed systematically, it is possible 
to observe the two amplitude switchings.
\end{abstract}



\maketitle

\section{Introduction}
\label{Introduction}

Vesicle dynamics has long been investigated both experimentally and 
theoretically~\cite{Huang,Farago,Schneider,Milner2,Onuki,Seki}. 
The decay rate of the thermally excited shape fluctuation provides information of 
vesicle elasticity. 
The knowledge of the microscopic relaxation can also be used to predict the macroscopic 
rheological property of the vesicle solution \cite{Seki}.

The vesicles are mostly non-equilibrium system. 
The external perturbation, be it thermal, electrical, 
sonication~\cite{sonication2,sonication3}, 
or flow~\cite{Diat,Le}, transforms the lamellar structure into the vesicles. 
In such situation, one often obtains mixture of uni-lamellar vesicle and 
multi-lamellar vesicles (MLV) of various sizes. 
However, in sharp contrast with the very detailed calculations and scattering experiments 
on the unilamellar vesicles, there are relatively few studies of the MLV 
dynamics in the literature. 
In this paper, we consider the two-bilayer vesicle which is the simplest MLV.
We calculate the dynamic relaxation rates, as well as the rheological response of the 
two-layer-vesicles. 
From the experimental point of view, it is hard to prepare vesicles with exactly 
two bilayers. 
Nonetheless, such a concrete calculation provides a clear picture of how the 
interaction between the two membranes affect the relaxation rates, as well as the 
squeezing (lubrication) flow which arises exclusively in MLV.
Since our method can be extend to vesicles with more than two bilayers, a more specific 
calculation can be performed similarly.

The squeezing dynamics of the sandwiched solvent between the two layers have been analyzed 
in two related problems. 
This relaxation process of the flat lamellar system was calculated by Brochard and 
de Gennes as the ``slip mode''~\cite{Brochard}, or later measured and called 
as ``baroclinic mode''~\cite{Nallet,Ramaswamy,Sigaud}. 
In the soap film system, the ``squeezing mode'' dispersion has been calculated and 
measured \cite{Fijnaut,Young}. 
In this work, we obtain the similar relaxation in which the solvent is squeezed 
between the two adjacent bilayers to relax the bilayer curvature energy and the 
mutual interaction energy between the two bilayers.   
As this mode arises only in MLV, we are interested in its dispersion relation and 
its coupling strength with the applied shear.
Due to the strong lubrication resistance, the squeezing mode is often the slowest 
relaxation mode.  
This makes the squeezing mode an important candidate for the vesicle rheology. 
However, not every relaxation mode is equally excited by the applied shear. 
Therefore whether the shear can drive the squeezing mode with a large amplitude is 
an important problem to consider.
In this work, we will try to build our understanding of the coupling strength through 
the concrete calculation.

Based on the analysis below, we find that the squeezing mode makes the dominant 
contribution to the complex viscosity for strongly interacting bilayers. 
Interestingly, when the gap between the two bilayers is smaller than a characteristic 
crossover gap, the fast undulation mode becomes the dominant mode for the complex 
viscosity. 
We find that the crossover gap gets very small when the bilayer interaction is strong.  
On the opposite limit where the bilayer has extremely weak interaction, the crossover 
gap becomes comparable to the inner vesicle radius.

Below in Sec.~\ref{sec:model}, we define our model. 
Section \ref{sec:elasticforce} discusses the elastic force. 
In Sec.~\ref{sec:normalforce} and Appendices, the flow resistance between the 
bilayers is solved. 
In Sec.~\ref{sec:spectrum}, we shall analyze the relaxation rate.
In Sec.~\ref{sec:viscoelasticity}, we consider the viscoelasticity of the 
dilute vesicle suspension.
Finally in Sec.~\ref{sec:summary}, we summarize and discuss our results.

\section{The model}
\label{sec:model}

\begin{figure}
\includegraphics[scale=0.4]{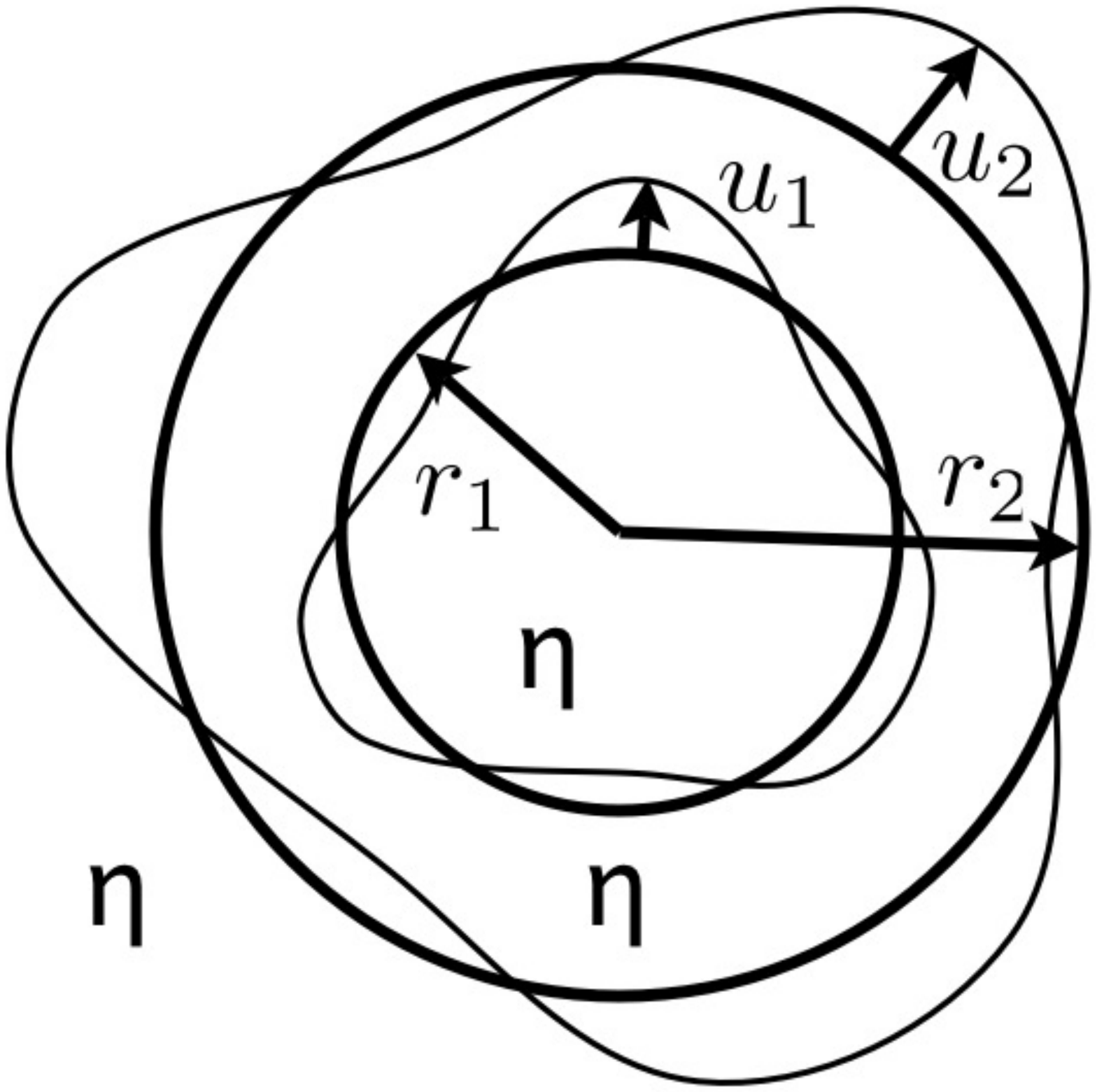}
\caption{The vesicle consists of two bilayers with the radius $r_1$ and $r_2$.
The surrounding solvent viscosity is $\eta$ everywhere.}
\label{fig1}
\end{figure}

We consider a two-layer-vesicle with two bilayers located at the spheres with 
mean radius $r_1$ and $r_2$ as shown in Fig.~\ref{fig1}. 
The deformation of the bilayer shapes are described by the radial layer 
displacements $u_1$ and $u_2$ relative to the two reference spheres respectively. 
Both of the bilayer membranes are surrounded by a solvent of viscosity $\eta$.
Let $a$ denote the area per molecule projected on the reference sphere, 
and $a_0$ the averaged projected area per molecule.
For each membrane, we define the dimensionless surface density $\phi_n$ 
($n=1,2$) given by the ratio $a_0/a$.
Notice that $\phi_n$ becomes unity at equilibrium. 
In a fluctuating vesicle, $\phi_n$ is not uniform in general.
In this work we propose the free energy of a two-layer-vesicle as 
\begin{align}
F &=\frac{(r_2^3- r_1^3)}{3}\frac{B }{2} \int 
\left(\frac{u_2 - u_1}{r_2-r_1} \right)^2 {\rm d} \Omega 
\nonumber \\
&+  \sum_{n=1}^2 \int 
\left[ \gamma_n +\frac{\kappa}{2}H_n^2 +\frac{E}{2}(\phi_n-1)^2 
\right] {\rm d} A_n, 
\label{F_cont}
\end{align}
where $B$ is the layer compression modulus, 
$\gamma_n$ the surface tension, 
$\kappa$ the bending modulus, and 
$E$ the area stretching/compression modulus. 
Here both $\kappa$ and $E$ are taken to be the same for the two membranes.
Moreover, ${\rm d} \Omega $ is the differential solid angle, and the surface 
area element is approximately given by 
${\rm d}A_n \approx [1+(\nabla_\perp u_n)^2/2] r_n^2 \, {\rm d} \Omega$.
The mean curvature $H_n$ is given by 
\begin{equation}
H_n \approx -\frac{2}{r_n}+\frac{2 u_n}{r_n^2}+\nabla_\perp^2 u_n,
\end{equation}
up to linear order in $u_n$.

Two comments should be made about the first term of Eq.~(\ref{F_cont}). 
The compression modulus $B$ should be a function of $r_1$ and $r_2$. 
In principle a microscopic statistical model for MLV should provide 
the functional form. 
Here we focus on the layer dynamics, therefore to build such a model 
is beyond the scope of this work. 
As a rough estimate for discussion, below in subsection \ref{sec:interaction}, 
we will use the $B$, which is calculated from the flat layers, and replace 
the distance $d$ between two layers by $r_2-r_1$. 
This approximation is justified for $d\ll r_2$, while it may deviate considerably
when $d$ is comparable to $r_2$.

The interaction term proposed here is proportional to $(u_2-u_1)^2$, which 
arises naturally at small $d$ from the square of the strain. 
At large $d$, the curvatures and the areas for the two bilayers are very 
different. 
Presumably a microscopic model may derive a more suitable weighted interaction 
energy, which is proportional to $[u_2-(r_1/r_2)^\beta u_1]^2$ with a weighting 
exponent $\beta$. 
Below we present the calculation without this extra weighting factor. 
Nonetheless, the calculation procedure is exactly the same for the case 
$\beta \not= 0$.

At the dilute phase boundary of the lamellar phase, the surface tension 
$\gamma_n$ should vanish. 
Inside the lamellar phase, the tension depends on the applied osmotic pressure. 
To simplify the discussion, we shall only consider the case where the surface 
tension vanishes $\gamma_n=0$. 
In fact, the calculation with finite surface tension can be carried out in 
the same way.  
The layer displacement $u_n$ is related to the radial velocity $v_r$ at $r=r_n$ by
\begin{equation}
\frac{\partial u_n}{\partial t}=v_r(r_n),
\end{equation}
where we neglect the solvent permeation.
The surface density obeys
\begin{equation}
\frac{\partial \phi_n}{\partial t}=-\frac{2 v_r(r_n)}{r_n}\phi_n-
\nabla_\perp \cdot [v_{\perp}(r_n) \phi_n],
\label{phi}
\end{equation}
where $\nabla_\perp$ is the two-dimensional (2D) surface derivative and 
$v_{\perp}(r_n)$ is the tangential velocity at $r=r_n$.
Here we have neglected the amphiphile exchange flux between the neighboring 
bilayer (the amphiphile permeation).

The flow field obeys the Stokes equation which is presented here as the form
\begin{equation}
\eta \nabla^2 \mathbf{v}-\nabla p=0,
\label{Stokes}
\end{equation}
where $\eta$ is the viscosity, and $p$ the pressure.
Here we write down the force balance conditions that are satisfied on the 
two layers.
The normal force balance is given by 
\begin{equation}
-\frac{\delta F}{\delta u_n} + \sigma_{rr} (r_n^+)-\sigma_{rr} (r_n^-)=0,
\label{normal_force}
\end{equation}
where $\sigma_{rr}=-p+2\eta \partial_rv_r$ ($r$ being the radial distance), 
and the superscripts $+$ and $-$ indicate that the stresses are evaluated at 
the exterior and interior of the bilayer, respectively.
On the other hand, the tangential force balance is given by 
\begin{equation}
-\frac{\delta F}{\delta \mathbf{x}_n} + \sigma_{\perp} (r_n^+)
-\sigma_{\perp} (r_n^-)=0,
\label{tangent_force_balance}
\end{equation}
where $\mathbf{x}_n$ is the 2D tangent displacement of the layers, and 
the 2D stress is defined as 
$\sigma_\perp=\hat{\theta} \sigma_{r\theta}+ \hat{\varphi}\sigma_{r\varphi}$
($\theta$ and $\varphi$ being the polar and azimuthal angles, respectively).

The viscoelastic response of a dilute two-layer-vesicle suspension can 
be calculated by considering the stress response to an external flow
at large distances from the vesicles
\begin{equation}
\mathbf{v}^\infty(\mathbf{r},t)=\Gamma \nabla 
\left[ r^2 Y_{20}(\theta,\varphi) \right] e^{i\omega t},
\label{v_shear}
\end{equation}
where $\Gamma$ is the strength of the elongational flow, 
$Y_{lm}(\theta,\varphi)$ are the spherical harmonics, 
and $\omega$ is the angular frequency.  
Each suspended vesicle contributes to the averaged stress. 
In the dilute solution, the effective complex viscosity is calculated 
as~\cite{Batchelor,Seki}
\begin{equation}
\eta^\ast=\eta \left( 1-\frac{p^{\rm II}_{20}}{4\eta \Gamma r_2^3 }
\phi_{\rm v} \right),
\label{effective_eta}
\end{equation}
where $\phi_{\rm v}=(4\pi/3)c r_2^3$ is the volume fraction occupied 
by the vesicles ($c$ is the number density of the vesicles), and 
$p^{\rm II}_{20}$ is the frequency dependent complex coefficient of 
the spherical harmonic expansion of the pressure (see 
Eqs.~(\ref{p_expansion}) and (\ref{pC}) later).
The effective viscosity can also be expressed in terms of the complex modulus 
as $G^{\ast}=i\omega \eta^{\ast}$.

\section{The elastic forces}
\label{sec:elasticforce}

\subsection{Force expressions}

The elastic forces are calculated by evaluating the derivatives 
$-\delta F/\delta u_n $ and $-\delta F/\delta \mathbf{x}_n$.
From Eq.~(\ref{phi}), we see that the bilayer tangential displacement and 
the radial displacement produce  the first order surface density 
perturbation as 
\begin{equation}
\delta \phi_n=-\frac{2u_n}{r_n}-\nabla_\perp \cdot \mathbf{x}_n. 
\label{delta_phi}
\end{equation}
The tangential elastic force is given by 
\begin{equation}
-\frac{\delta F}{\delta \mathbf{x}_n}=\nabla_\perp (\delta \gamma_n),
\label{tangent_force}
\end{equation} 
where the tension perturbation $\delta \gamma_n$ is induced by the surface 
density perturbation, i.e., $\delta \gamma_n = - E (\phi_n-1)$.
Notice that the 2D gradient has the usual component form; 
\begin{equation}
\nabla_\perp (\delta \gamma_n)= \hat{\theta} \frac{1}{r} 
\frac{\partial(\delta \gamma_n)}{\partial \theta} + 
\hat{\varphi} \frac{1}{r\sin \theta} 
\frac{\partial (\delta \gamma_n)}{\partial \varphi}.
\end{equation}

Up to linear order in $u_n$ and $\delta \gamma_n$, the normal force on bilayer 
1 is given by 
\begin{align}
-\frac{\delta F}{\delta u_1}& =\frac{B(r_2^3-r_1^3)}{3 (r_2-r_1)^2}
\frac{u_2-u_1}{r_1^2} -\frac{2(\gamma_1+\delta \gamma_1)}{r_1} \nonumber \\
& + (\gamma_1-\kappa \nabla_\perp^2 )
\left( \nabla_\perp^2+\frac{2}{r_1^2} \right) u_1,
\end{align}
whereas that for bilayer $2$ is 
\begin{align}
- \frac{\delta F}{\delta u_2}& =- \frac{B(r_2^3-r_1^3)}{3 (r_2-r_1)^2}
\frac{u_2-u_1}{r_2^2} -\frac{2(\gamma_2+\delta \gamma_2)}{r_2}
\nonumber \\
& +(\gamma_2-\kappa \nabla_\perp^2 )
\left( \nabla_\perp^2+\frac{2}{r_2^2} \right) u_2.
\end{align}
In the large stretching modulus limit ($E \rightarrow \infty$), the tension 
perturbations $\delta \gamma_n$ become Lagrange multipliers, so that they 
ensure that the right hand side of Eq.~(\ref{delta_phi}) vanishes. 
By using Eqs.~(\ref{tangent_force_balance}) and (\ref{tangent_force}), 
the values of $\delta \gamma_n$ are determined from the viscous stress 
$\sigma_\perp$ (see Appendix C).

\subsection{The bilayer interactions}
\label{sec:interaction}

There are several interactions which contribute to the layer compression 
modulus $B$. 
In this subsection, we indicate their physical origins and give the simplest 
formulae to describe them.
For charged bilayers, the electrostatic interaction, together with the 
counter-ion and co-ion entropy produce the free energy per 
area~\cite{Israelachvili,Safran,Russel}
\begin{equation}
V_{\rm e}=\frac{64 C_{\rm s} k_{\rm B}T}{\kappa_{\rm D}}
\tanh^2\left( \frac{q\psi}{4k_{\rm B}T} \right) e^{-\kappa_{\rm D} d},
\end{equation}
where $C_{\rm s}$ is the salt concentration, 
$q \psi$ the potential energy of the counter-ion $q$ at the surface, 
$\kappa_{\rm D}$ the inverse of the Debye screening length, 
$d = r_2-r_1$ the distance between the two surfaces, 
and $k_{\rm B}T$ the thermal energy.

The van der Waals attraction potential per unit area can be calculated by 
summing the dipoles to obtain
\begin{equation}
V_{\rm vdW} \simeq -\frac{A}{12 \pi}\left[ \frac{1}{(d-\delta)^2} 
+\frac{1}{(d+\delta)^2}-\frac{2}{d^2} \right],
\end{equation}
where $A$ is the Hamaker constant, and $\delta$ is the membrane thickness. 
A more complicated Lifschitz theory calculation can provide a more accurate 
description~\cite{Israelachvili,Safran,Russel,Parseigian}. 
The sum of these two interactions consists the standard DLVO theory. 
When the electrostatic repulsion and van der Waals attraction stabilize MLV,
$B$ can be evaluated from
\begin{equation}
B = d \frac{\partial^2 (V_{\rm e}+V_{\rm vdW})}{\partial d^2}.  
\label{DLVO}
\end{equation}

For a flexible bilayer where the undulation entropy depends strongly on the 
membrane separation $d$, Helfrich estimated the free energy per area and 
$B$ by a self-consistent argument to obtain~\cite{Helfrich1,Safran} 
\begin{equation}
B=c_0\frac{(k_{\rm B}T)^2}{\kappa d^3},
\label{Helfrich}
\end{equation}
where $c_0$ is an order unity numerical constant. 
In the following examples below, we use $c_0=36/\pi^2$.
When the van der Waals interaction becomes important, a more elaborate 
calculation is required to combine the undulation entropy and the van der 
Waals attraction~\cite{Lipowsky}.

\section{The normal force balance}
\label{sec:normalforce}

Using the spherical harmonic expansion, the Stokes equation can be solved 
in terms of the radial functions which are simple polynomial of the radius. 
The detailed calculation is presented in the Appendix A. 
In general, the solution depends on the boundary velocities, which are 
the velocities of the membranes and the applied external shear. 
In the usual case where the bilayers have a large stretching modulus $E$,
we can simplify the discussion by considering the large stretching modulus 
limit $E \rightarrow \infty$. 
In this limit, the surface density approaches unity, and the combinations 
$E (\phi_n-1)$ become the Lagrange multipliers to ensure that surface 
densities are constant. 
Hence the constant $E$ can be eliminated.
The detailed calculation is presented in Appendix B.
At the boundaries, this limit also turns the tangential velocity into the 
function of the normal velocity.
This means that the normal stresses $\sigma_{rr}(r_n^+)$ and $\sigma_{rr}(r_n^-)$, 
as well as the tension perturbation $\delta \gamma_n$ are all proportional 
to the bilayer normal velocities and the external shear. 
The detailed calculation is presented in Appendix C.

The bilayer displacements and velocities are expanded as the sum of the 
spherical harmonics $Y_{lm}(\theta,\varphi)$ times the time varying amplitudes,
i.e., $u_n(\theta, \varphi)=u_n Y_{lm}(\theta,\varphi)$ and 
$v_r(r_n, \theta, \varphi)=v_n Y_{lm}(\theta,\varphi)$.
Without cluttering the notation with the angular quantum numbers $l$ and $m$, 
hereafter we use $u_n$ and $v_n$ to express the time varying amplitudes 
for an arbitrary set of $(l, m)$.
After performing some calculations to express $\sigma_{rr}$ and $\delta \gamma_n$ 
by $v_n$ (see Eqs.~(\ref{C1}), (\ref{C5}), (\ref{pA}), (\ref{pC}) and (\ref{pB})), 
we find that the normal force balance Eq.~(\ref{normal_force}) at $r_1$ and $r_2$ 
can be written as
\begin{equation}
\mathbf{E} \cdot
\left(
\begin{array}{c}
 r_1^2 u_1 \\
  r_2^2 u_2
 \end{array}
\right) 
+\mathbf{D} \cdot
\left(
\begin{array}{c}
 r_1^2 v_1 \\
  r_2^2 v_2
\end{array}
\right) 
= 20 \eta \Gamma e^{i\omega t}\delta_{l2}\delta_{m0} \hat{\mathbf{e}}_2 ,
\label{linear_equation}
\end{equation}
where $\hat{\mathbf{e}}_2=(0,1)$.
Here we prefer to use the variables $r_1^2 u_1$ and $r_2^2 u_2 $ which make 
the matrices $\mathbf{E}$ and $\mathbf{D}$ symmetric. 
With this variable choice, the free energy per solid angle is then the 
quadratic form of $\mathbf{E}$.  The components of the matrix 
$\mathbf{E}$ are given by 
\begin{align}
E_{11}&=\frac{B (1-\rho^3)}{3r_2^3(1-\rho )^2\rho^4}
+\frac{(\gamma_1 r_2^2\rho^2+\kappa \hat L^2 ) (\hat L^2-2 )}{r_2^6\rho^6}, 
\nonumber \\
E_{12}&=-\frac{B (1-\rho^3)}{3r_2^3 (1-\rho )^2\rho^2 },
\nonumber \\
E_{21}&= E_{12},
\nonumber \\
E_{22}&=\frac{B (1-\rho^3)}{3r_2^3 (1-\rho )^2}
+\frac{(\gamma_2 r_2^2+\kappa \hat L^2 )(\hat L^2-2 )}{r_2^6}, 
\end{align}
where have defined $\rho = r_1/r_2 $ ($\le 1$) and
\begin{equation}
{\hat L^2} = -\frac{1}{\sin \theta} \frac{\partial}{\partial \theta}
\left( \sin \theta \frac{\partial}{\partial \theta} \right)
-\frac{1}{\sin^2 \theta } \frac{\partial^2}{\partial \varphi^2}.
\end{equation}
Whereas the components of the matrix $\mathbf{D}$ are 
\begin{align}
D_{11}&= \frac{\eta r_2^{4l+1} (2l+1)}{ G_0 (l^2+l)} \times \nonumber \\ 
&  \left[ -(l+1)^2(4l^2-1)\rho^{2l+1} \right. \nonumber \\ 
&  +(l^2-1)(2 l-1)(2l+3)\rho^{2l-1} \nonumber \\
&  +(l-1)^2(4 l^2+8l+3)\rho^{2l-3} \nonumber \\
&  \left. +(8l^2+8l-4)\rho^{-2} \right],
\nonumber \\
D_{12}&=\frac{\eta r_2^{4l+1} (4l+2)}{G_0  (l^2+l)} \times \nonumber \\
&  \left[ (2l^3+5l^2+l-2)(\rho^{3l+2} -\rho^{l-1}) \right.
\nonumber \\ 
& \left. +(2l^3+l^2-3l)(\rho^{l+1}-\rho^{3l})\right],
\nonumber \\
D_{21}&=D_{12},
\nonumber \\
D_{22}&=\frac{\eta r_2^{4l+1} (2l+1)}{ G_0  (l^2+l)} \times \nonumber \\
&  \left[ -(l+2)^2(4l^2-1)\rho^{2l+4} \right. \nonumber \\ 
&  +2l(l+2)(2l-1)(2l+3)\rho^{2l+2} \nonumber \\
&   +l^2 (4l^2 + 8 l +3)\rho^{2l} \nonumber \\
&  \left. +(8l^2+8l-4) \rho  \right],
\end{align}
where 
\begin{align}
G_0 & = r_2^{4l+4} \left[4\rho+4 \rho^{4l+3} 
-  (2l+1)^2\rho^{2l+4} \right. \nonumber \\
& \left. - (6-8l-8l^2)\rho^{2l+2}-(2l+1)^2 \rho^{2l} \right].
\end{align}
Notice that the components of $\mathbf{D}$ are the product of $\eta/r_2^3$ 
and the functions of $\rho$ and $l$.

When the two bilayers are well separated, i.e., $r_1 \ll r_2$, they are not 
hydrodynamically coupled. 
In this limit, $D_{12}$ and $D_{21}$ become small, and we recover the 
isolated vesicle damping given by~\cite{Milner2,Onuki,Seki}
\begin{equation}
D_{nn} = \frac{\eta(2l+1)(2l^2+2l-1)}{r_n^3l(l+1)}.
\end{equation}

For the more general bilayer interaction which is proportional 
to $B [u_2-(r_1/r_2)^\beta u_1]^2$, the similar calculation will give 
rise to an extra factor $\rho^{2\beta}$ for the $B$ term in $E_{11}$, 
and an extra factor $\rho^\beta$ for the $B$ terms in $E_{12}$ and $E_{21}$. 
The form of $E_{22}$ is not affected.

\section{Relaxation spectrum}
\label{sec:spectrum}

The relaxation rates, denoted as $\Omega_j$ ($j=1,2$), are the eigenvalues 
of the matrix $\mathbf{D}^{-1}\cdot \mathbf{E}$.
The normal force balance condition Eq.~(\ref{normal_force}) on the two bilayers 
gives the eigenvector equation  
\begin{equation}
\mathbf{D}^{-1} \cdot \mathbf{E} \cdot \mathbf{d}_j^{\rm R} 
=\Omega_j \mathbf{d}_j^{\rm R},
\label{ei1}
\end{equation}
where $\mathbf{d}_j^{\rm R}$ are the (non-orthogonal) right-eigenvectors. 
To work with the orthogonal eigenvectors, here we define the symmetric decay 
rate matrix
\begin{equation}
\mathbf{M} = 
\mathbf{D}^{-1/2} \cdot \mathbf{E} \cdot \mathbf{D}^{-1/2}, 
\end{equation}
which has the (normalized) eigenvectors $\mathbf{d}_j$
\begin{equation}
\mathbf{M} \cdot \mathbf{d}_j =\Omega_j \mathbf{d}_j.
\end{equation}
Here the eigenvalues are the same as in Eq.~(\ref{ei1}), and 
$\mathbf{d}_j$ is proportional to $\mathbf{D}^{1/2} \mathbf{d}_j^{\rm R}$.
We can decompose the decay rate matrix as
\begin{equation}
\mathbf{M}=\Omega_1 \mathbf{d}_1 \mathbf{d}_1 
+ \Omega_2 \mathbf{d}_2 \mathbf{d}_2, 
\label{decompose}
\end{equation} 
which will be used in the later discussion.
Hereafter the faster and the slower rates are denoted by $\Omega_1$ and 
$\Omega_2$, respectively.

The full expression of the decay rates are straightforward, but complicated. 
Hence we shall obtain some simplified expressions to gain some understanding 
on the nature of the relaxation. We first discuss the fast mode $\Omega_1$.
When $\rho=1$, the fast undulation mode has a well-known dispersion 
relation~\cite{Schneider,Milner2,Seki}
\begin{equation}
\Omega_0=\frac{2\kappa (l-1)l^2(l+1)^2(l+2)}{\eta r_2^3 (2l+1)(2l^2+2l-1)},
\label{omega0}
\end{equation}
where the bending modulus is doubled as there are two bilayers.
For $\rho$ slightly less than unity, a series expansion can be made if desired. 
For $\rho \ll 1$, the decay rate can be approximated by 
\begin{equation}
\Omega_1 \approx \frac{E_{11}}{D_{11}}.
\label{omega1}
\end{equation}
This corresponds to a simple picture that the fast mode consists mainly the inner 
bilayer relaxation, whereas the outer bilayer does not move much.

Next we consider the slow mode $\Omega_2$. 
When $\rho \approx 1$, the slow mode can be approximately obtained by the series 
expansion of $1-\rho$. 
 Based on the relation 
$\Omega_1+\Omega_2={\rm tr}(\mathbf{D}^{-1} \cdot \mathbf{E})$, in which 
$\Omega_2$ only starts from the second order term, we get 
$\Omega_1\simeq {\rm tr}(\mathbf{D}^{-1} \cdot \mathbf{E})$ 
for the zeroth and the first order terms. 
Because the product $\Omega_1\Omega_2$ is given by
${\rm det} (\mathbf{D}^{-1} \cdot \mathbf{E})$, we use the leading two 
terms of ${\rm det} (\mathbf{D}^{-1} \cdot \mathbf{E})/\Omega_1 $ to obtain 
\begin{align}
\Omega_2 & \approx \frac{B l (l+1) }{12 \eta}  (1-\rho )^2 (2-\rho)
\nonumber \\
& + \frac{ \kappa(l-1) l^2 (l+1)^2(l+2)}{24 \eta r_2^3} (1-\rho)^3.
\label{B2}
\end{align}
The leading term $(B/12 \eta)(1-\rho )^2l(l+1)$  is similar to
the slip mode of the planar smectic, which has the decay rate 
$(B/12\eta) d^2q_\perp^2$, where $q_\perp$ is the wave vector projected on 
the bilayer plane~\cite{Brochard}.
This expression is also analogous to the squeezing mode dispersion obtained for 
soap film~\cite{Fijnaut,Young}.
In the other extreme of $\rho = 0$, the slow mode has the dispersion such that 
\begin{equation}
\Omega_2 \approx \frac{\left[B+3(l-1)l(l+1)(l+2))\kappa /r_2^3 \right] l(l+1)}
{3 \eta (2l+1)(2l^2+2l-1)}.
\label{B2a}
\end{equation}
If we set the shear perturbation as $l=2$, the numerator contains a factor 
$B+72 \kappa/r_2^3$. 
Hence the dimensionless parameter $Br_2^3/\kappa$ becomes important when its 
value is much larger than 72.

We now discuss another approximate expression for the slow mode valid for 
the whole range of $\rho$.  
Since the sum of the two decay rates 
$\Omega_1+\Omega_2={\rm tr}(\mathbf{D}^{-1} \cdot \mathbf{E})$
is dominated by the fast mode for any $\rho$, the slow mode can be approximated 
by the ratio 
\begin{align}
\Omega_2 & \approx \frac{\det (\mathbf{D}^{-1} \cdot \mathbf{E})}
{{\rm tr}(\mathbf{D}^{-1} \cdot \mathbf{E})} \nonumber \\
& = \frac{E_{11}E_{22}-E_{12}E_{21}}
{E_{11}D_{22}-E_{12}D_{21}+E_{22}D_{11}-E_{21}D_{12}}.
\end{align}
When the inner bilayer 1 relaxes fast, 
 one can apply the adiabatic approximation to the fast relaxing inner 
membrane. 
The approximated expression has even a simpler denominator given by
\begin{equation}
\Omega_2 \approx \frac{E_{22}-E_{21}(E_{12}/E_{11})}{D_{22}-D_{21}(E_{12}/E_{11})}.
\label{slow_mode}
\end{equation}

\begin{figure}
\includegraphics[scale=0.4]{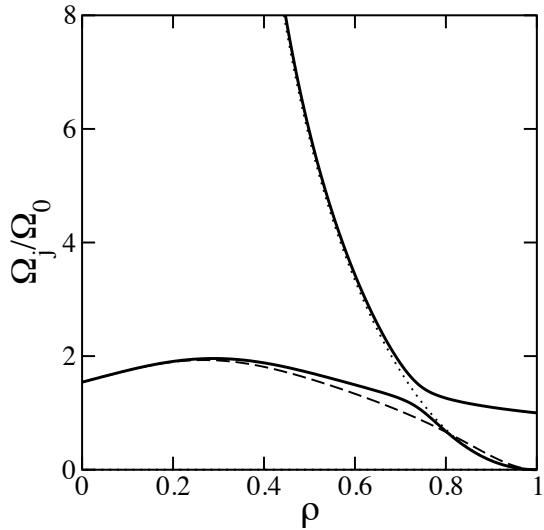}
\caption{The scaled relaxation rate $\Omega_j/\Omega_0$ as a function of the 
dimensionless size ratio $\rho=r_1/r_2$ between the two layers. 
Here $\Omega_0$ is the vesicle relaxation rate with the rigidity $2\kappa$
given by Eq.~(\ref{omega0}). 
The two solid lines represent the two vesicle relaxation rates 
$\Omega_1$ and $\Omega_2$ obtained numerically.
The dotted line is an approximation for the fast mode given by Eq.~(\ref{omega1}).
The dashed line represents the approximate slow relaxation rate $\Omega_2$ 
given by Eq.~(\ref{slow_mode}). 
In this plot, we set $B=600$ J/m$^3$, 
$\kappa =k_{\rm B}T=4 \times 10^{-21}$ J, 
$\eta=10^{-3}$ Pa$\cdot$s, 
$r_2=10^{-7}$ m, $l=2$, so that $B r_2^3/\kappa =150$.}
\label{fig2}
\end{figure}

In Fig.~\ref{fig2}, we plot the decay rates as a function of $\rho=r_1/r_2$ for $l=2$,  keeping a constant $B$.
The two solid lines represent the numerical calculated decay rates.
The upper one is the fast mode $\Omega_1$ which has the limit $\Omega_0$ at $\rho=1$.
It coincides with the approximation Eq.~(\ref{omega1}) (dotted line) for $\rho < 0.6$.
The lower solid line represents the slow mode $\Omega_2$.
The approximate slow rate Eq.~(\ref{slow_mode}) follows qualitatively the exact 
value of $\Omega_2$ for the full range of $\rho$. 
It also coincides with  Eq.~(\ref{B2a}) in the limit of $\rho=0$. 
Incidentally the inner layer relaxation Eq.~(\ref{omega1}) also fits the slow 
mode $\Omega_2$ at $\rho \approx 1$.

Fig.~\ref{fig2} is useful to illustrate the nature of the relaxation for the 
fast and slow modes. 
In a real system, $B$ is a function of $r_1 $ and $r_2$ (or $\rho$ and $r_2$), 
which is hard to be kept as a constant. 
One will make a plot with a function for $B$ which is suitable for the specific 
MLV system. 
If one uses the flat layer formulas described in subsection \ref{sec:interaction} 
as approximations, caution should be kept for the accuracy at small $\rho$. 
Given an accurate function for $B$, together with the proper weighting factor 
(to combine $\rho^{2\beta}$ and $\rho^\beta$ factors in $\bf E$), the above decay 
rate approximations should work better.

\section{Viscoelasticity}
\label{sec:viscoelasticity}

\begin{figure}
\includegraphics[scale=0.4]{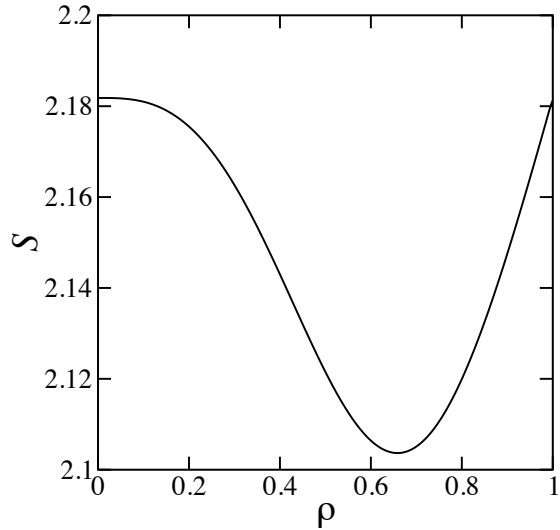}
\caption{The shear coupling strength $S$ defined by Eq.~(\ref{shear}) 
as a function of the dimensionless size ratio $\rho$ between the two layers.}
\label{fig3}
\end{figure}

We now consider two-layer-vesicles under the external oscillatory shear flow. 
Rearranging Eq.~(\ref{effective_eta}) together with Eq.~(\ref{pC}), we obtain
\begin{equation}
\frac{\eta^{\ast}/\eta-1}{\phi_{\rm v}}=\frac{5}{2}-
\frac{v_2}{r_2 \Gamma e^{i\omega t}}.
\label{high_shear_eta}
\end{equation}
It is apparent that the dilute hard sphere limit is recovered when $v_2 = 0$.
By putting $u_n=v_n/i\omega$ and solving Eq.~(\ref{linear_equation}) for $v_2$, 
we obtain
\begin{equation}
\frac{\eta^{\ast}/\eta-1}{\phi_{\rm v}}=\frac{5}{2}-
i\omega\frac{20 \eta}{r_2^3} \hat{\mathbf{e}}_2 \cdot 
({\bf E}+i\omega {\bf D} )^{-1} \cdot \hat{\mathbf{e}}_2,
\label{high_shear_eta2}
\end{equation}
where one should set $l=2$ for the matrixes $\mathbf{E}$ and $\mathbf{D}$. 
Here $\hat{\mathbf{e}}_2$ appears twice, because the shear directly affects 
the bilayer 2 through Eq.~(\ref{linear_equation}), and the velocity of the 
bilayer 2 carries  the stress contribution of the vesicle according to 
Eqs.~(\ref{effective_eta}) and (\ref{pC}).

Considering the high-frequency limit, we now define the dimensionless shear 
coupling strength $S$ and the shear deformation unit vector $\hat{\mathbf{s}}$ by 
\begin{equation}
S^{1/2} \hat{\mathbf{s}}=\left( \frac{20 \eta}{r_2^3} \right)^{1/2}
\mathbf{D}^{-1/2} \cdot \hat{\mathbf{e}}_2.
\label{shear}
\end{equation}
Notice that $S$ and $\hat{\mathbf{s}}$ only depend on the ratio $\rho$ as long as 
we fix to $l=2$.
As shown in Fig.~\ref{fig3}, $S$ is weakly depended on $\rho$ because its 
value varies only between 2.1 and 2.19.  
Hence the right hand side of Eq.~(\ref{high_shear_eta2}) varies between 0.31 and 0.4, 
whereas the hard sphere case gives 2.5.
This means that, at high frequencies, the coupling of a vesicle to the flow is 
weaker compared to that of a hard sphere.

\begin{figure}
\includegraphics[scale=0.4]{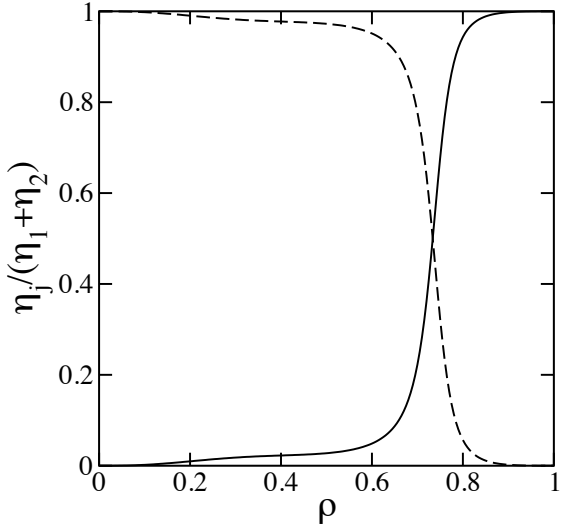}
\caption{The relative viscosity strengths defined by Eq.~(\ref{eta_n})
as a function of the dimensionless size ratio $\rho$ between the two layers.
The solid and the dashed lines corresponds to $\eta_1$ and $\eta_2$, 
respectively.
We chose the value $B r_2^3/\kappa=150$.
The crossing of the two modes occurs at $\rho^{\ast} \approx 0.733$.}
\label{fig4}
\end{figure}

The high frequency viscosity asymptotically approaches
\begin{equation}
\eta^\infty = \eta \left[ 1+\left(\frac{5}{2}- S \right) \phi_{\rm v}\right]
=\eta^0-\eta S \phi_{\rm v},  
\end{equation}
where $\eta^0=\eta [1 +(5/2)\phi_{\rm v}]$.
Using the eigenmode decomposition 
$\mathbf{I}=\mathbf{d}_1 \mathbf{d}_1 + \mathbf{d}_2 \mathbf{d}_2$, we can 
separate the viscosity contributions from each mode as
\begin{equation}
\eta_j = S \eta \left( \hat{\mathbf{s}} \cdot \mathbf{d}_j \right)^2,
\label{eta_n}
\end{equation} 
so that we have
\begin{equation}
\eta S=\eta_1+\eta_2.
\label{sum}
\end{equation}
Then the full expression for the complex viscosity is given by 
\begin{equation}
\eta^{\ast}(\omega)=\eta^0-\phi_{\rm v} \sum_{j=1}^2 
\eta_j \frac{i\omega}{i\omega+\Omega_j}. 
\end{equation} 
The relative strength of the two modes are shown in Fig.~\ref{fig4}.
Alternatively (but equivalently), the complex modulus can be expressed as
\begin{equation}
G^{\ast}(\omega)= i\omega \eta^\infty + \phi_{\rm v} 
\sum_{j=1}^2 G_j \frac{\omega^2+i\omega \Omega_j}{\omega^2+\Omega_j^2},
\label{G*}
\end{equation}
where the two modes have the positive amplitudes $G_j = \Omega_j \eta_j$.

\begin{figure}
\includegraphics[scale=0.4]{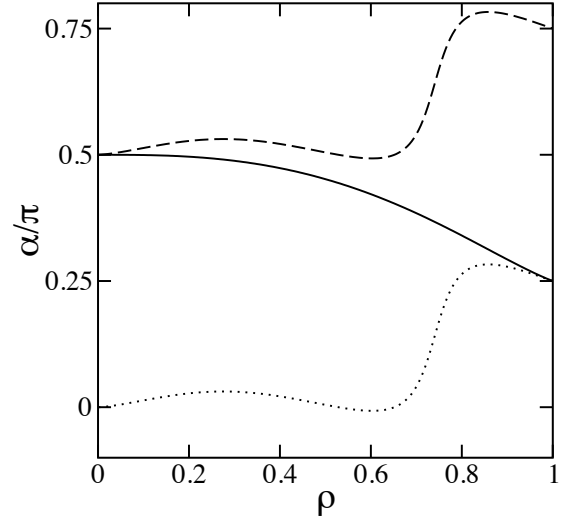}
\caption{The polar angle $\alpha$ (divided by $\pi$) of various 2D vectors 
as a function of the dimensionless size ratio $\rho$ between the two layers.
The solid line represents $\hat{\mathbf{s}}$, while the dotted and the
dashed lines represent $\mathbf{d}_1$ and $\mathbf{d}_2$, respectively.
We chose the value $Br_2^3/\kappa=150$.}
\label{fig5}
\end{figure}

Even for different bilayer interaction strength $B$, we always find that the 
slow mode has the larger viscosity amplitude at $\rho \approx 0$, and smaller 
one at $\rho \approx 1$. 
The latter is reasonable because the shear perturb the two bilayers similarly 
for $\rho \approx 1$, where $\bf \hat s$ is along the $(1,1)$-direction. 
In terms of the polar angle $\alpha$ between $\bf \hat s$ and the first axes  on the 
$(r_1^2 u_1, r_2^2 u_2)$-plane, the $(1,1)$-direction corresponds to 
$\alpha=\pi/4$.
Notice that this direction also corresponds to the undulation eigenmode 
direction.
Since the sum of the two viscosity amplitudes is roughly a constant as 
shown in Eq.~(\ref{sum}), the slow squeezing mode takes a small viscosity 
amplitude $\eta_2$ for $\rho \approx 1$.
In the other limit of $\rho \approx 0$, the shear mainly perturbs the outer 
layer.
In this limit, we have ${\bf \hat s} = (0,1)$ or $\alpha=\pi/2$,
and the slow mode is due to the outer layer relaxation.
As a result, $\eta_2$ becomes the dominant contribution for $\rho \approx 0$. 
In Fig.~\ref{fig5}, we plot the angle $\alpha$ as a function of $\rho$. 
The shear vector $\bf \hat s$  always coincides with the fast mode at $\rho=1$ 
and the slow mode at $\rho=0$. 
Therefore the mode switching always takes place between $0 \le \rho \le 1$. 
 It is also worth mentioning that Figs.~(\ref{fig4}) and (\ref{fig5}), 
like Fig.~(\ref{fig2}), are plotted by assuming a constant $B$. 
These plots are used to analyze the behavior of the mode amplitudes. 
The more physical plots will be the ones using a suitable function for $B$.

\begin{figure}
\includegraphics[scale=0.4]{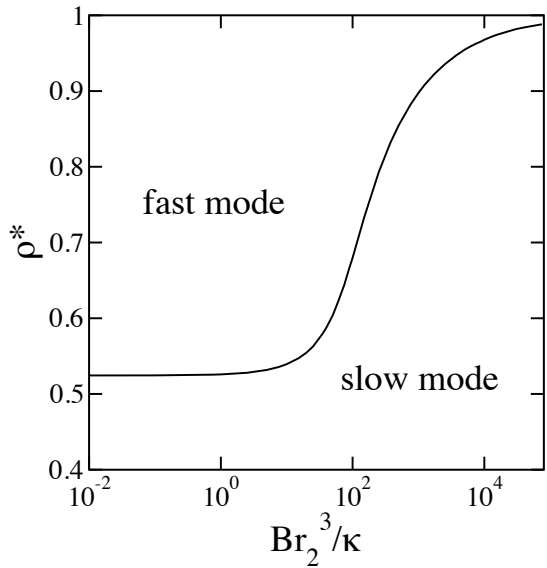}
\caption{The crossover size ratio $\rho^{\ast}$ as a function of the dimensionless
bilayer interaction $B r_2^3/\kappa$. 
For $\rho > \rho^{\ast}$, the fast mode is the dominant contribution to the 
viscosity, while the slower relaxation is dominant for $\rho < \rho^{\ast}$.}
\label{fig6}
\end{figure}

Here we define a crossover size ratio $\rho^{\ast}$ at which the two modes have the 
same viscosity amplitude, i.e., $\eta_1=\eta_2$.
In Fig.~\ref{fig6}, we plot the calculated $\rho^{\ast}$ as a function of 
$Br_2^3/\kappa$. 
When $Br_2^3/\kappa \ll 1$, the crossover happens at $\rho^{\ast} \approx 0.52$.
This means that for non-interacting case $B=0$, the crossover happens when 
$d=r_2-r_1$ is roughly the same as $r_1$. 
When $Br_2^3/\kappa \gg 1$, on the other hand, $\rho^{\ast}$ approaches unity.
As mentioned in Eq.~(\ref{B2a}), the parameter $Br_2^3/\kappa $ will have a  
noticeable effect only when it is greater than 72. 
At the lower right corner of Fig.~\ref{fig6}, where the layer interaction is 
strong and the layer separation is not too small, the slow mode has the 
dominant viscosity contribution. 
At the upper left corner where the layer interaction is relatively weak, the 
fast mode is more excited as compared to the slow mode.

\section{Summary and Discussion}
\label{sec:summary}

In summary, we have calculated the slow relaxation rates and the viscoelasticity 
of a dilute two-layer-vesicle solution.
We have found the following points: 
(i) At small gap $\rho \approx 1$, the slowest mode is the squeezing mode. 
The undulation mode appears to be faster.
(ii) When the inner bilayer radius is small $\rho \approx 0$, the slow mode 
becomes the relaxation of the outer bilayer, and the faster mode is the relaxation 
of the inner bilayer. 
(iii) When one decreases $\rho$, the relaxation spectrum changes from (i) to (ii).
(iv) For the complex viscosity, the low frequency viscosity approaches the 
hard sphere limit. 
The high frequency viscosity increment is between 12--16 \% of the hard sphere 
viscosity increment. 
The difference between the two limits comes from the contributions of the two modes.
(v) At small gap $\rho > \rho^{\ast}$, the slow (squeezing) mode has a small viscosity 
amplitude, while the fast (undulation) mode has a large viscosity amplitude. 
For large gap $\rho < \rho^{\ast}$, on the other hand, the slow mode has the dominant 
viscosity amplitude. 
(vi) The crossover size ratio $\rho^{\ast}$ depends on the interaction between the 
two bilayers. 
As $B r_2^3/\kappa$ is increased, $\rho^{\ast}$ increases from 0.52 toward unity.

\begin{figure}
\includegraphics[scale=0.4]{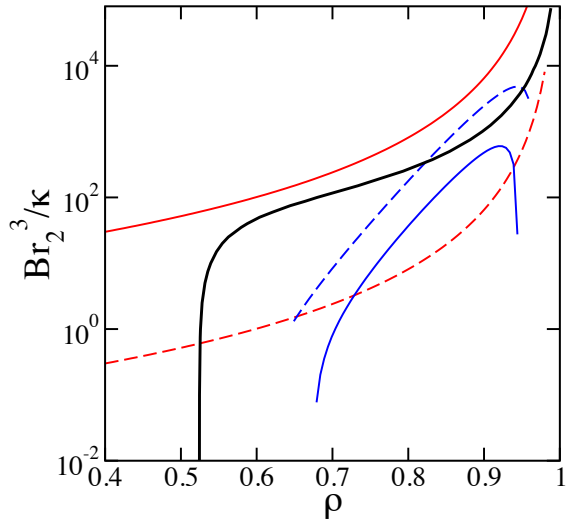}
\caption{(color online) 
The dimensionless bilayer interaction $B r_2^3/\kappa$ as a function of the 
size ratio $\rho$. 
The crossover ratio $\rho^{\ast}$ in Fig.~\ref{fig6} is plotted by the solid 
black line.
The blue lines are from the DLVO theory Eq.~(\ref{DLVO}) with 
$C_{\rm s}=0.01$ M, $\kappa_{\rm D}=3 \times 10^{-8}$ m, $A=10^{-21}$ J, 
$\delta=3 \times 10^{-9}$ m, $k_{\rm B}T=4 \times 10^{-21}$ J.
The surface potentials are $12$ mV and $24$ mV for the blue solid and dashed lines, 
respectively.
The red lines are the Helfrich repulsion Eq.~(\ref{Helfrich}) with $c_0=36/\pi^2$,  
$r_2=10^{-7}$ m.  
The red solid and dashed lines are for 
$\kappa=0.75k_{\rm B}T=3 \times 10^{-21}$ J and 
$\kappa=7.5k_{\rm B}T=3 \times 10^{-20}$ J, respectively.}
\label{fig7}
\end{figure}

We have determined the crossover ration $\rho^\ast$ in Fig.~\ref{fig6}. 
The actual bilayer interaction strength $B$ depends on the separation 
$d=r_2(1-\rho)$. 
For the qualitative discussion at non-small $\rho$, we use the flat 
layer results in Sec.~\ref{sec:interaction}.
In Fig.~\ref{fig7} we present both information in the $(\rho ,Br_2^3/\kappa)$-plot
to compare a series of systems with the same outer bilayer size $r_2$ but with 
different size ratios $\rho$.
As one varies $\rho$, the dimensionless bilayer interaction strength 
$Br_2^3/\kappa$ changes according to the DLVO theory Eq.~(\ref{DLVO}) (blue lines) 
or the Helfrich repulsion Eq.~(\ref{Helfrich}) (red lines). 
If such lines happen to cross the line $\rho^\ast$ (black line), 
one expects that the two viscosity amplitudes change their relative magnitudes. 
We have indicated such scenarios by the dashed lines.

As shown in the blue dashed line in Fig.~\ref{fig7}, an electrostatically stabilized 
system may get large $B$ by having a high surface charge density. 
At small gap, the van der Waals attraction may also lower $B$, causing an 
interesting multiple crossing.
This means that the viscosity amplitudes switch their magnitudes more than once. 
For sterically stabilized system (red lines), we find that the dimensionless 
interaction parameter depends strongly on the bending rigidity as  
$Br_2^3/\kappa \sim \kappa^{-2}$ (see Eq.~(\ref{Helfrich})). 
For a soft surfactant bilayer of $\kappa=0.75 k_{\rm B}T$, we find that the slow mode 
always dominates the viscosity. 
Whereas for a lipid bilayer whose bending rigidity $\kappa=7.5 k_{\rm B}T$ is one 
order of magnitude larger, the red dashed line indicates that the squeezing mode is 
not much excited by shear when $\rho > 0.526$.

In this paper, we have only presented the calculation for two-layer-vesicles. 
For MLV with more than two layers, the calculation can be performed in the same 
procedure, but with a greater algebraic complexity. 
For MLV with $N$ layers, we expect that there are $N$ relaxation modes.
When the gap is small ($r_{N-1}/r_N \approx 1$), we expect that the majority of 
the $N$ relaxations to bear some resemblance to the squeezing mode or the spherical 
version of the ``slip mode''~\cite{Brochard}. 
For larger gap, on the other hand, the relaxations of the $N$ layers may decouple 
from each other. 
As for the viscoelasticity, we speculate that the viscosity amplitudes of the 
squeezing modes are small for weakly interacting MLV.
Neutral or weakly charged lipid MLV should be an interesting system to investigate 
in this direction.

In polymer rheology calculation, one may consider a step strain for $t \ge 0$, 
where the time relaxation of the stress gives the relaxation modulus $G(t)$.
This quantity can be further converted to the complex modulus $G^{\ast}(\omega )$ 
by the Fourier transform.  
Right after the step strain, one often assumes that the polymer deforms in an affine way.  
Do we implicitly use the affine deformation approximation for the MLV rheology? 
A step strain for vesicle solution will induce a short but fast bilayer movement, 
where the $\mathbf{D}$ terms dominate the left hand side of  Eq.~(\ref{linear_equation}). 
Therefore  the initial layer displacements will be proportional to 
$\mathbf{D}^{-1} \cdot \hat{\mathbf{e}}_2  \propto (-D_{12}/D_{11},1)$ which is 
compatible with the incompressible constraints (see Eq.~(\ref{2Dincompressible})) 
at both layers. 
In the small gap limit, $\mathbf{D}^{-1} \cdot \hat{\mathbf{e}}_2$ behaves like 
$(\rho^2 ,1)$.
Compared with the affine deformation $u_n \propto r_n$, or in terms of our chosen 
variables $(r_1^2u_1, r_2^2u_2) \propto (\rho^3, 1)$, it is clear that our 
theory does not use the affine deformation approximation.

\begin{acknowledgments}

We thank M.\ Cates and S.\ Fujii for many useful discussions.
We also acknowledge the support from the National Science Council of Taiwan and 
the Center of Theoretical Physics of National Taiwan University.
SK also acknowledges the supported by Grant-in-Aid for Scientific Research (grant 
No.\ 24540439) from the MEXT of Japan, and the JSPS Core-to-Core Program 
{\it ``International research network for non-equilibrium dynamics of soft matter''}.
\end{acknowledgments}

\appendix

\section{Solution of Stokes equation}

For the incompressible solenoidal flow, the velocity can be expressed as
\begin{equation}
\mathbf{v}= \nabla \times (\nabla \psi \times \mathbf{r})  
+\nabla \times (\zeta \mathbf{r}),
\end{equation}
where the scalar functions $\psi$ and $\zeta$ are the defining function for 
the poloidal and toroidal flow fields, respectively. 
Note that our definition of the defining function differs by a factor $r$ 
compared with the ones in the book by Chandrasekhar~\cite{Chandrasekhar}.

For an incompressible fluid, the divergence of the  pressure gradient vanishes,
i.e., 
\begin{equation}
\nabla^2 p=0.
\label{p}
\end{equation}
Therefore the pressure gradient is also a solenoidal field, and can be described 
by the above decomposition. 
Since the toroidal part can be written as $(\nabla \zeta) \times \mathbf{r}$,
the condition $\partial_\theta (\partial_\varphi p)=\partial_\varphi (\partial_\theta p)$ 
requires that $\partial_\theta (-\sin \theta \ \partial_\theta \zeta )=
\partial_\varphi (\partial_\varphi \zeta /\sin \theta )$ or ${\hat L^2}\zeta =0$,
where 
\begin{equation}
{\hat L^2} = -\frac{1}{\sin \theta} \frac{\partial}{\partial \theta}
\left( \sin \theta \frac{\partial}{\partial \theta} \right)
-\frac{1}{\sin^2 \theta } \frac{\partial^2}{\partial \varphi^2}.
\end{equation}
Hence the pressure gradient is only poloidal, and written as
\begin{align}
\nabla p &=\nabla \times (\nabla \Psi \times \mathbf{r}) \nonumber \\ 
&= \hat{r} \frac{\hat L^2 \Psi}{r} 
+ \hat{\theta} \frac{1}{r} \frac{\partial}{\partial \theta}
\frac{\partial (r \Psi)}{\partial r} 
+ \hat{\varphi} \frac{1}{r \sin \theta} \frac{\partial}{\partial \varphi}
\frac{\partial (r \Psi)}{\partial r},
\label{component}
\end{align}
where $\Psi$ is the defining function of $\nabla p$.
Comparing the $\hat{\theta}$ and $\hat{\varphi}$ directions, we can set 
\begin{equation}
p=\frac{\partial(r \Psi)}{\partial r},
\label{p2}
\end{equation}
for these two directions. 
The Laplace equation for the pressure Eq.~(\ref{p}) then implies that
\begin{equation}
\frac{1}{r^2} \frac{\partial}{\partial r} \left( r^2 
\frac{\partial^2 (r \Psi)}{\partial r^2} \right) 
- \frac{\hat L^2}{r^2} \frac{\partial (r \Psi)}{\partial r}=0.
\label{p_components}
\end{equation}
Therefore the radial component of the gradient pressure can be either 
expressed as $\hat L^2 \Psi/r$ or $\partial_r^2 (r \Psi )$. 
The latter expression is consistent with the identification Eq.~(\ref{p2}).

For the velocity field, we will drop the toroidal part by setting $\zeta=0$.
This is justified because the tangential force field is a surface gradient and
drives only the poloidal flow. 
To obtain the defining function $\psi$,  we take the curl of Eq.~(\ref{Stokes}) 
to get $\nabla \times \nabla^2 \mathbf{v}=0$. 
Since $\nabla^2 \mathbf{v}=-\nabla \times \nabla \times \mathbf{v}$ holds for 
incompressible flow, we have 
\begin{equation} 
\nabla \times \nabla \times \nabla \times \nabla \times [\nabla \times  
(\psi  \mathbf{r})]=0,
\end{equation}
where in the square bracket, an equivalent form of $\nabla \psi \times \mathbf{r}$
is used.
As detailed in Ref.~\cite{Chandrasekhar}, each curl will switch poloidal 
and toroidal parts. 
The double curl will preserve the type and modify the defining function 
by $-\nabla^2$.  
The curl of Eq.~(\ref{Stokes}) becomes 
$\nabla \times (\mathbf{r} \nabla^4 \psi)=0$ or
\begin{equation}
\nabla^4 \psi=0.
\label{psi}
\end{equation}

To find the coupling between the pressure $p$ and the velocity poloidal 
function $\psi$ in Eq.~(\ref{Stokes}), we rewrite 
$-\nabla^2 \mathbf{v}=\nabla \times \nabla \times \mathbf{v}
=\nabla \times [\nabla (-\nabla^2 \psi )\times \mathbf{r}]$.
Then Eq.~(\ref{Stokes}) is satisfied when
\begin{equation}
-\Psi+\eta\nabla^2 \psi=0.
\nonumber
\end{equation}
Here we prefer to use $p$ instead of $\Psi$. 
Operating the above equation by $\hat L^2/r$ and substituting the 
combination $\hat L^2 \Psi/r$ by $\partial_r p$, we obtain
\begin{equation}
r \frac{\partial p}{\partial r} =\eta {\hat L^2} \nabla^2 \psi.
\label{ppsi}
\end{equation}

Using the standard method of separation of variables, the general solution of 
Eqs.~(\ref{p}), (\ref{psi}) and (\ref{ppsi}) are
\begin{equation}
p=p_0+\sum_{lm}\left( p^{\rm I}_{lm} r^l +p^{\rm II}_{lm}r^{-l-1} \right) Y_{lm}, 
\label{p_expansion}
\end{equation}
\begin{align}
\psi &=\sum_{lm}\left( \psi^{\rm I}_{lm}r^l +\psi^{\rm II}_{lm}r^{-l-1} 
\right) Y_{lm} \nonumber \\ 
&+\sum_{lm}\left( \frac{p^{\rm I}_{lm}r^{l+2}}{\eta (2l+2)(2l+3)}  
+\frac{p^{\rm II}_{lm}r^{-l+1}}{2\eta l(2l-1)} \right) Y_{lm},
\end{align}
where $Y_{lm}(\theta,\varphi)$ are the spherical harmonics.

\section{The surface incompressible limit}

The surface incompressibility condition also constrains the velocity field as 
\begin{equation}
\frac{2v_r(r_n)}{r}+\nabla_\perp \cdot \mathbf{v}_\perp(r_n)=0
\label{2Dincompressible}
\end{equation}
at both $r=r_1$ and $r_2$. 
Because the fluid is incompressible, i.e., $\nabla \cdot \mathbf{v}=0$, 
Eq.~(\ref{2Dincompressible}) can also be expressed as the equivalent form
\begin{equation}
\frac{\partial v_r(r_n)}{\partial r}=0,
\label{2Dincompressible2}
\end{equation}
at the bilayers. 
We prefer this condition because it simplifies the calculation.

We now consider the relaxation rates of a small perturbation described by 
the spherical harmonics $Y_{lm}(\theta,\varphi)$. 
To simplify the notation, we drop the subscript ``$lm$'' of the coefficients 
$\psi_{lm}^{\rm I}$, $\psi_{lm}^{\rm II}$, $p_{lm}^{\rm I}$, and $p_{lm}^{\rm II}$. 
For the region within the bilayer 1 ($0\le r \le r_1 $), we set 
$\psi_{lm}^{\rm II}=0$ and $p_{lm}^{\rm II}=0$ to drop the functions which are 
singular at $r=0$. 
We denote $\psi_{lm}^{\rm I}$ and $p_{lm}^{\rm I}$ as $\psi_{\rm A}^{\rm I}$ and 
$p_{\rm A}^{\rm I}$, respectively. 
The 2D incompressibility condition Eq.~(\ref{2Dincompressible2}) relates the 
two remaining coefficients as
\begin{equation}
\psi_{\rm A}^{\rm I}=\frac{r_1^2}{2\eta (2l+3)(l-1)}p_{\rm A}^{\rm I}.
\end{equation}
In terms of the radial velocity amplitude $v_1$, we can further express 
the pressure coefficient as
\begin{equation}
p_{\rm A}^{\rm I}=-\frac{\eta(2l+3)(l-1)r_1^{-l-1}}{l} v_1.
\label{pA}
\end{equation}

For the region exterior to the bilayer 2 ($r_2 \le r < \infty $), the coefficients $\psi_{lm}^{\rm I}$ are zero, except 
$\psi_{20}^{\rm I}$, which needs to be chosen to give the far flow Eq.~(\ref{v_shear}). 
Comparing the radial component $v_r$ from Eq.~(\ref{v_shear}) and $\hat L^2 \psi/r$, 
we set $\psi_{20}^{\rm I}=\Gamma/3$.
We also set $p_{lm}^{\rm I}=0$ so that $p$ does not diverge at 
$r \rightarrow \infty$, and denote $\psi_{lm}^{\rm II}$ and $p_{lm}^{\rm II}$ as 
$\psi_{\rm C}^{\rm II}$ and $p_{\rm C}^{\rm II}$, respectively.
Then the 2D incompressibility condition Eq.~(\ref{2Dincompressible2}) relates 
the two remaining coefficients as
\begin{equation}
\psi_{\rm C}^{\rm II} =-\frac{r_2^2}{2\eta (2l-1)(l+2)}p_{\rm C}^{\rm II}  
+ \frac{1}{12} r_2^{5} \Gamma \delta_{l2}\delta_{m0}.
\end{equation}
Using the radial velocity amplitude $v_2$, we can express the pressure coefficient as
\begin{equation}
p_{\rm C}^{\rm II} =\frac{\eta(2l-1)(l+2)r_2^{l}}{l+1} v_2  
-10 \eta r_2^3 \Gamma \delta_{l2}\delta_{m0}.
\label{pC}
\end{equation}

For the region between the two bilayers ($r_1 \le r < r_2 $), the expressions are 
more complex. 
The incompressibility Eq.~(\ref{2Dincompressible2}), evaluated at $r_1$ and 
$r_2$, provides two conditions between the four coefficients 
\begin{align}
\psi_{\rm B}^{\rm I} &=F_{11}p_{\rm B}^{\rm I}+F_{12}p_{\rm B}^{\rm II}, \nonumber \\
\psi_{\rm B}^{\rm II}&=F_{21}p_{\rm B}^{\rm I}+F_{22}p_{\rm B}^{\rm II}, 
\end{align}
where
\begin{align}
F_{11}&=-\frac{r_2^{2l+3}-r_1^{2l+3}}
{2\eta (2l+3)(l-1)\left(r_2^{2l+1}-r_1^{2l+1}\right)}, 
\nonumber \\
F_{12}&=\frac{r_2^{2}-r_1^{2}}{2\eta (2l-1)(l-1)\left(r_2^{2l+1}-r_1^{2l+1}\right)}, 
\nonumber \\
F_{21}&=-\frac{r_1^{2l+1}r_2^{2l+1}(r_2^{2}-r_1^{2})}
{2\eta (2l+3)(l+2) \left(r_2^{2l+1}-r_1^{2l+1}\right)}, \nonumber \\
F_{22}&=-\frac{r_1^2 r_2^{2l+1}-r_2^2 r_1^{2l+1}}
{2\eta (2l-1)(l+2)\left(r_2^{2l+1}-r_1^{2l+1}\right)}. 
\end{align}
We prefer to use the variables $v_1$ and $v_2$ instead of $p_{\rm B}^{\rm I}$ 
and $p_{\rm B}^{\rm II}$. 
Then the pressure coefficients can be expressed as  
\begin{align}
p_{\rm B}^{\rm I} &=G_{11}v_1+G_{12}v_2, \nonumber \\
p_{\rm B}^{\rm II}&=G_{21}v_1+G_{22}v_2, 
\label{pB}
\end{align}
where
\begin{align}
G_{11} &= \frac{\eta}{l G_0}  \left\{ (-8l^2-4l+12)r_1^{3l+2}r_2 \right.
\nonumber \\ 
&- \left. (4l+6)r_1^{l+1}r_2^{2l}[l(2l+1)r_1^2-(l+2)(2l-1)r_2^2] \right\},
\nonumber \\
G_{12}&=  \frac{\eta}{l G_0} \left\{ (-8l^2-4l+12)r_1 r_2^{3l+2} \right.
\nonumber \\ 
&- \left. (6l+4)r_1^{2l} r_2^{l+1}[l(2l+1)r_2^2-(l+2)(2l-1)r_1^2] \right\}, 
\nonumber \\
G_{21}&=  \frac{(4l-2)\eta}{(l+1)G_0} \left\{ -(2l^2+3l+1)r_1^{3l+4} r_2^{2l} \right.
\nonumber \\ 
&+  \left. (2l^2+l-3)r_1^{3l+2} r_2^{2l+2}+ (2l+4)r_1^{l+1} r_2^{4l+3} \right\},
\nonumber \\
G_{22}&= \frac{(4l-2)\eta}{(l+1)G_0} \left\{ -(2l^2+3l+1)r_2^{3l+4} r_1^{2l} \right.
\nonumber \\ 
&+  \left. (2l^2+l-3)r_2^{3l+2} r_1^{2l+2}+ (2l+4)r_2^{l+1} r_1^{4l+3} \right\},
\end{align}
with 
\begin{align}
G_0 & = r_2^{4l+4} \left[4\rho+4 \rho^{4l+3} 
-  (2l+1)^2\rho^{2l+4} \right. \nonumber \\
& \left. - (6-8l-8l^2)\rho^{2l+2}-(2l+1)^2 \rho^{2l} \right].
\end{align}
When the two bilayer are well separated ($r_1 \ll r_2$), both $G_{11}$ and 
$G_{22}$ become small.
In this limit, $G_{12}$ becomes the coefficient of Eq.~(\ref{pA}) with 
$r_1$ replaced by $r_2$, whereas $G_{21}$ becomes the coefficient of 
Eq.~(\ref{pC}) (without the $\Gamma$ term) with $r_2$ replaced by $r_1$.

\section{The stress and the tension perturbation}

The radial component of the normal stress appears in Eq.~(\ref{normal_force}). 
From its component form $\sigma_{rr}=-p+2\eta \partial_r v_r$ and the 2D 
incompressibility condition Eq.~(\ref{2Dincompressible2}), it is just the 
negative pressure. 
Therefore the stress differences at the two bilayers are
\begin{align}
\sigma_{rr}(r_1^+)-\sigma_{rr}(r_1^-) &=
\sum_{lm}\left[ (p_{\rm A}^{\rm I}-p_{\rm B}^{\rm I})r_1^l 
-p_{\rm B}^{\rm II}r_1^{-l-1} \right] Y_{lm}, \nonumber \\
\sigma_{rr}(r_2^+)-\sigma_{rr}(r_2^-) &=
\sum_{lm}\left[ p_{\rm B}^{\rm I} r_2^l 
+(p_{\rm B}^{\rm II}-p_{\rm C}^{\rm II})r_2^{-l-1} \right] Y_{lm}.
\label{C1}
\end{align}

The large bilayer stretching modulus $E$ suppresses the surface density 
perturbation. 
For the slow relaxation, we will take the limit $E \rightarrow \infty$ to 
eliminate the parameter $E$. 
Within this limit, both $\phi_1$ and $\phi_2$ approach unity, so that the 
tension perturbation $\delta \gamma_1$  and $\delta \gamma_2$ become the 
Lagrange multipliers. 
Then the values of the Lagrange multipliers are determined by 
Eqs.~(\ref{tangent_force_balance}) and (\ref{tangent_force}) instead of 
their original definitions, as discussed below.

In Eq.~(\ref{tangent_force_balance}), we are interested in the difference 
between the tangential stress across the bilayer. 
We express the tangential stress as a function of $\psi$:
\begin{equation}
\sigma_\perp=\hat{\theta} \sigma_{r\theta} + \hat{\varphi} \sigma_{r\varphi} 
= \eta \nabla_\perp \left[ r \frac{\partial^2 \psi}{\partial r^2}
+(\hat{L}^2-2) \frac{\psi}{r} \right]. 
\end{equation}
We now replace $\partial_r^2\psi$ using Eq.~(\ref{ppsi}), and limit our 
discussion to $l \neq 0$ mode.
The velocity field $\nabla \times (\nabla \psi \times \mathbf{r})$ has the 
components similar to Eq.~(\ref{component}). 
We can use the $r$ component $v_r=\hat L^2\psi /r$ to eliminate $\psi$ as 
$(r/\hat L^2)v_r$ so that 
\begin{equation}
\sigma_\perp=\nabla_\perp \left[ \frac{r^2}{\hat L^2}
\frac{\partial p}{\partial r} 
+\eta \left(2-\frac{4}{\hat L^2} \right) v_r
-2\eta \frac{r}{\hat{L}^2} \frac{\partial v_r}{\partial r} \right], 
\end{equation}
where $\hat{L}^2 \rightarrow l(l+1)$ for spherical harmonics with nonzero $l$.
Because $v_r$ is continuous across the bilayer and $\partial_r v_r$ vanishes 
on the bilayer, only the first term can be different across the bilayer.
This is the only important term for the tension perturbation in 
Eqs.~(\ref{tangent_force_balance}) and (\ref{tangent_force}):
\begin{equation}
\delta \gamma_n  = -\frac{r_n^2}{\hat{L}^2}
\left( \frac{\partial p}{\partial r} \right)_{r_n^+} +
\frac{r_n^2}{\hat{L}^2}
\left( \frac{\partial p}{\partial r} \right)_{r_n^-}.
\end{equation}
We then obtain
\begin{align}
\delta \gamma_1&= \sum_{lm}
\left(  \frac{p_{\rm A}^{\rm I}-p_{\rm B}^{\rm I}}{l+1} r_1^{2l+1}
+\frac{p_{\rm B}^{\rm II}}{l} r_1^{-l} \right)Y_{lm}, \nonumber \\
\delta \gamma_2&= \sum_{lm}
\left( \frac{p_{\rm B}^{\rm I}}{l+1} r_2^{2l-1}
+\frac{p_{\rm C}^{\rm II}-p_{\rm B}^{\rm II}}{l} r_2^{-l}\right)Y_{lm}. 
\label{C5}
\end{align}


\end{document}